\documentclass[twocolumn,prl,amsmath,amssymb,floatfix]{revtex4}

\newcommand{\be}{\begin{equation}}
\newcommand{\ee}{\end{equation}}
\newcommand{\bn}{\begin{displaymath}}
\newcommand{\en}{\end{displaymath}}
\newcommand{\bs}{\begin{eqnarray}}
\newcommand{\es}{\end{eqnarray}}

\newcommand{\Pf}{\text{Pf}\;}
\newcommand{\stk}[1]{\stackrel{#1}{\longrightarrow}}

\def\figwidth{0.9\linewidth}
\usepackage[final]{graphicx}

\begin{document}

<

\title{Logarithmic corrections to the Ising model free energy on lattices with conical singularities}
\date{June 16, 2003}

\author{Ruben Costa-Santos
\footnote{e-mail R.A.Costa-Santos@phys.uu.nl}}
\affiliation{Spinoza Institute, Utrecht University, Leuvenlaan 4, 3584 CE Utrecht}

\preprint{SPIN 03-13 ITP-UU 03-23}

\begin{abstract}

The free energy  of a two-dimensional system at criticality has in general an universal part proportional the logarithm of the system size. This term was shown by Cardy and Peschel to be related to the curvature of the system, with smooth metrics and singular points contributing in distinct ways. In this paper we present a numerical study of the effect, for the Ising model on lattices with various topologies from the sphere to genus two surfaces. The nature of this term for specific lattice models is an open problem  because the distinction between the two kinds of contributions involves an interchange of the limit of singular curvature with the thermodynamic limit. For all the lattices studied we found precise agreement with the conformal field theory prediction for conical singularities.

\end{abstract}

\maketitle

\section{Introduction}\label{section0}

The free energy of a two-dimensional system, of characteristic size $L$,  is expected from general arguments \cite{cardy2} to have, at criticality and for fixed shape, a large $L$ expansion of the form
\be
      F=  f_0  L^2 + f_b L + C \ln{L} + D +o(1)\label{fexp}.
\ee

While the bulk  free energy $f_0$ and the boundary free energy  $ f_b$  depend on the system details the coefficients $C$ and $D$ are universal. 
The scale independent term $D$ is known for strip geometries to be simply related to the theory central charge \cite{cardy} and for lattices without boundaries to be a modular invariant quantity \cite{me1,me2} determined by the lattice shape.

The coefficient of the logarithmic term $C$, from the conformal field theory point of view,  was shown by Cardy and Peschel \cite{cardy2} to be, for a system with a smooth metric and boundary,
\be
     C= -\frac{1}{6} \,c\, \chi \label{c1}
\ee
where  $\chi$ is the Euler characteristic of the system and  $c$ the central charge. 
This result can be understood  as the integral, over the all system, of the trace anomaly which is  proportional to the local curvature. However in the presence of  scale invariant geometric shapes, as boundary wedges or a bulk conical singularities,  the expression above fails to hold. For instance, for a conical singularity with deficit angle $\epsilon=2\pi-\theta$, where $\theta$ is the angle spanned by the cone, the contribution to the free energy is \cite{cardy2}     
\be
     C_\theta= \frac{c\,\theta}{24\pi} \left(1- (2\pi/\theta)^2 \right)  \label{c2}
\ee
and not a term proportional to $-\epsilon$ as one would expect from a delta function singularity on the curvature.

This  result is surprising because it signals that, concerning the free energy, a conical singularity cannot be considered as a limiting case of a smooth metric. Apparently the thermodynamic limit, with the ultraviolet cutoff going to zero, does not commute with the limit of taking  the curvature at a point to infinity. It is then natural to ask, following reference \cite{cardy2}, which of the expressions (\ref{c1}) or (\ref{c2}), if any, describes the logarithmic correction $C$ for a model defined on a regular lattice. 

In this paper we consider this problem for the Ising model defined on square lattices, without boundaries, on various topologies from the sphere to genus two surfaces.  The Ising model on lattices with the topology of the sphere has been  studied previously \cite{gonzalez0,hlang} but the coefficient  $C$ was obtained only for a restricted class of hexagonal lattices \cite{gonzalez}. 

\begin{figure}[b]
\includegraphics[width=\figwidth]{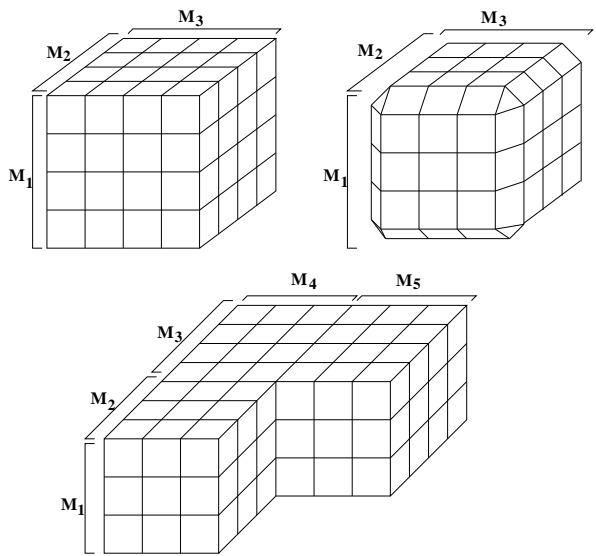}
\caption{The lattices with the topology of the sphere: the cubic shaped lattice, its dual and the L-shaped lattice. Each lattice is characterized by a set of integer dimensions $M_i$.}\label{fig1}
\end{figure}

\begin{figure}[t]
\includegraphics[width=\figwidth]{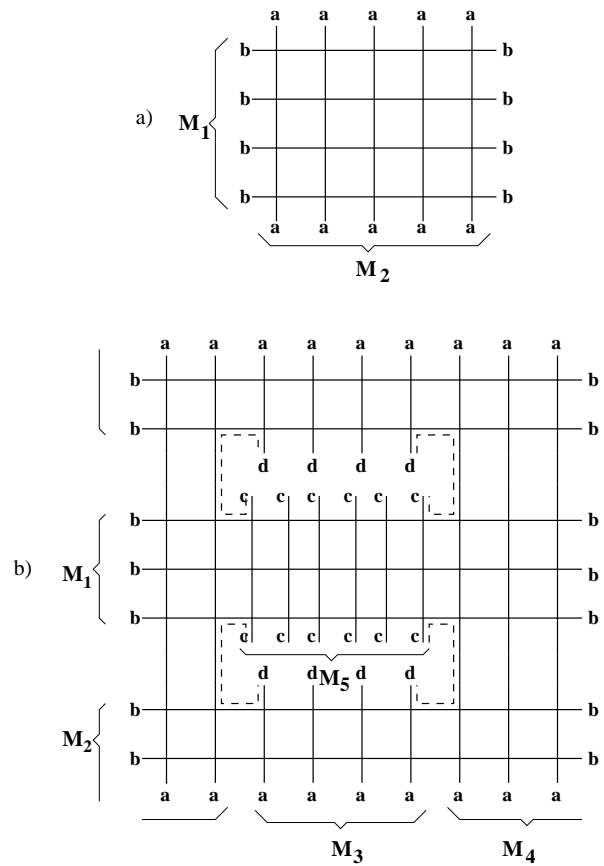}
\caption{The higher genus lattices: a) the usual $M_1$ rows by $M_2$ columns  toroidal lattice and b) a genus two lattice characterized by five integer sizes $(M_i:i=1,\ldots,5)$. The identifications of edges at the boundaries is given by the letters. The genus two lattice can be seen  as a torus with an additional handle on the bulk.}
\label{fig2}
\end{figure}

We consider three lattices  with the topology of the sphere;  the cube lattice, its dual and the L-shaped lattice shown in Fig. \ref{fig1}. We consider also the genus two lattice  shown in Fig. \ref{fig2}, that can be visualized as a torus with an additional handle \cite{me1}. All these lattices are locally flat, equivalent to the regular square lattice, with the exception of a few corners that can be seen as conical singularities.
For the cube and the L-shaped lattice these are the vertices where three or five squared faces meet. In the dual cube lattice the singularities are located on the triangular faces, that share edges with only three squared faces by opposition to four. For the genus two lattice of Fig. \ref{fig2} the singularities correspond to the two octagonal faces, shown in dashed line, that have eight neighboring squared faces. 

Closed form expressions of the Ising model free energy are known for toroidal \cite{kauf}, cylindrical  \cite{mccoy} and some non-orientable \cite{brankov,luwu2} geometries, where translational invariance permits the use of Fourier transforms. 
For the lattices considered in this paper, such an analytical treatment is not possible. The  free energy  will be  evaluate numerically, using the Kasteleyn method, for a series of lattices with fixed shape and increasing size.

The logarithmic correction  $C$  that we obtain for the various lattices considered, corresponds very accurately to the sum over the lattice singularities of the conical singularity contributions $C_\theta$ as given by Eq. (\ref{c2}).

\section{Evaluation of the free energy}\label{section1}

The Ising model partition function, on a two dimensional lattice, can be expressed using the Kasteleyn method \cite{mccoy,kast2,russ} in terms of the Pfaffians of certain antisymmetric adjacency matrices. 
The formalism has a combinatorial interpretation in terms of the high temperature polygon expansion of partition function. It can alternatively by seen as  graphical prescription to express the  Ising model partition function as a Grassman integral. 
The resulting Pfaffians, for an arbitrary lattice, cannot in general be evaluated in closed form but provide a good numerical tool to calculate the free energy for lattices with reasonably large size. In this section we will review general aspects of the Kasteleyn formalism  needed  to evaluate the partition function for the lattices of Fig. \ref{fig1} and \ref{fig2}. For additional details and proofs of the method we direct the reader to references  \cite{kast3,russ,me1}.

Consider a lattice, or more properly a graph,  $G$ drawn without superposition of edges on a surface of genus $g$, where the genus is the number of handles in the surface related with the Euler characteristic by $\chi= 2-2g$. The Ising model on such a lattice is defined by assigning a coupling constant to each edge connecting two lattice vertices and placing an Ising spin at each vertex.

\begin{figure}[t]
\includegraphics[width=\figwidth]{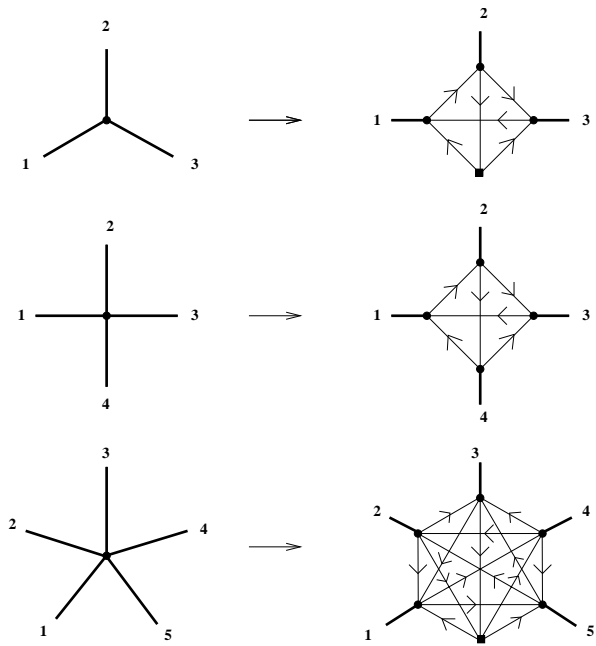}
\caption{The vertex decorations used for vertices with three, four and five incoming edges.}
\label{fig3}
\end{figure}

Following the Kasteleyn method we start by decorating the lattice vertices  with a decoration graph, meaning that each vertex on the lattice $G$ is  replaced by a cluster of vertices.  In the Grassman formulation, each vertex of the decoration corresponds to a Grassman variable associated with the original lattice site. The resulting decorated lattice will be denoted by $G_d$.
Our choice of decoration for vertices with coordination number three, four and five is shown in Fig. \ref{fig3}. The order by which the exterior edges connect to the decoration polygon is not important and the decoration graphs can be rotated.  

The edges of the decorated lattice $G_d$ are assigned  weights: $1$ to the internal edges of the decorations and a weight $w=\tanh{K}$ to the edges inherited from the lattice $G$, with $K$ being the Ising coupling constant on that edge in units of $K_B T$. 
The edges of $G_d$ need also to be oriented, assigned a direction that can be represented  graphically by an arrow. The internal edges of the decorations  have already an orientation given in Fig. \ref{fig3}. The remaining edges, inherited from the original lattice $G$, are oriented according to the Kasteleyn rule: in such a way that all lattice faces have an odd number of clockwise oriented edges. This condition  can be understood as a consistency condition for the  Grassman integral description, corresponding to the concept of  spin structure \cite{me1} in a spinor formalism.

\begin{figure}[b]
\includegraphics[width=\figwidth]{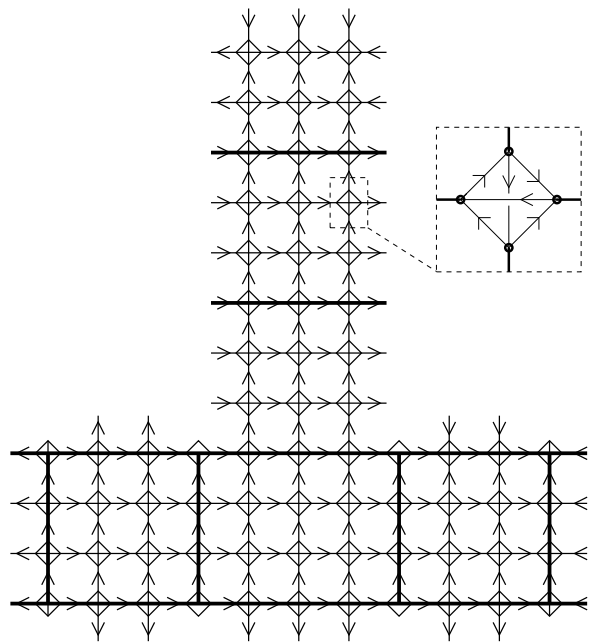}
\caption{An orientation of the decorated cube lattice that verifies the Kasteleyn rule. The edges along the boundary are identified to form a cube with dimension $4\times 5\times 4$. At each lattice site there is a  four point decoration as shown in the inset.}
\label{fig31}
\end{figure}

An example of a decorated lattice orientation, that satisfies the Kasteleyn condition, is shown in Fig. \ref{fig31}, for the cube lattice with dimensions $(M_1,M_2,M_3)=(4,5,4)$.
For a given lattice $G_d$ there are many different edge orientations that verify this condition. In fact given a Kasteleyn orientation we can generate a distinct Kasteleyn orientation by simply reversing the orientations (or arrows) on all the edges meeting at a given vertex.  Two lattice orientations related by such an arrow reversal, or a series of arrow reversals, are said to be equivalent. 

It can be showed \cite{me1,kast3,russ} that for a genus $g$ lattice there are precisely $4^g$ un-equivalent Kasteleyn orientations,  matching the number of spin structures on the corresponding continuum description.  This number is determined by the dimensions of the first homology group of the embedding surface. If  $ a_i, b_i$ with $i=1,\ldots,g$ are loops in the surface forming a canonical basis of the first homology group then the $4^g$ un-equivalent orientation can be generated from an initial Kasteleyn orientation by reversing the orientation of the edges crossed by the a choice of such loops. 
 
Given a Kasteleyn orientation of the decorated graph $G_d$ and a labeling of its vertices with an integer from 1 to ${\cal M}$, we define the corresponding adjacency matrix  $A$ as a ${\cal M}\times {\cal M}$ matrix with entries $A_{ij}$ that vanish if vertices $i$ and $j$ are not connected by an edge, and take a value $\pm z$ if  vertices $i$ and $j$ are connected by an edge of weight $z$, the sign being determined by the edge orientation. Schematically 
\be
       A_{ij}= \left\{
              \begin{array}{rl}
              z &  \textstyle{\text{ if    }} i \stk{z} j  \\
             -z &  \textstyle\text{{ if    }} j \stk{z} i  \\
              0 & \textstyle\text{{ otherwise}}
               \end{array}   \right. .
\ee

The Ising model partition function on the lattice $G$ can then be expressed as
\be
  Z^{\text{Ising}}(K) = 2^{N_V} (\cosh{K})^{N_V} \;\frac{1}{2^g}\; \sum_{i=1}^{4^g} \alpha_i \, \Pf A_i(K)  \label{zz}
\ee
were $N_V$ and $N_E$ are the number of vertices and edges in $G$ and we assume that all edges have the same coupling constant $K$. The sum runs over representatives of the  $4^g$  un-equivalent Kasteleyn orientations. The $\alpha_i$ take values $\pm 1$ and are completely determined by the arrow parity of the non-trivial topology loops $ a_i, b_i$  along lattice edges. Modulo these signs the Pfaffians above do not depend on which orientation is chosen from each equivalence class or on the choice of the labeling of  vertices. 

For translational invariant lattices, such as the toroidal square lattice, the Pfaffians of these adjacency matrices can be evaluated in a closed form. For the lattices shown in Figs. \ref{fig1} and \ref{fig2} such an analytic treatment is not possible and we are forced to resort to numerical evaluations of the $\Pf A_i$, the Pfaffian of an antisymmetric matrix being the square root of its determinant.

\section{Lattices with the topology of the sphere} \label{section2}

For the genus zero lattices shown in Fig. \ref{fig1} the partition function is expressed, according to Eq. (\ref{zz}), by the Pfaffian of a single adjacency matrix. This corresponds to single Pfaffian used by Kasteleyn \cite{kast1} to evaluate dimer coverings on the rectangular lattice with edges, notice that a spherical lattice can always be deformed and drawn in the plane without superposition of edges.

We evaluated the free energy for sequences of lattices with fixed shape and increasing size, by taking the integer dimensions characterizing the lattice $M_i$ to be of the form $M_i=m_i\, L$ with fixed $m_i$ and increasing $L$. The Ising coupling constant was fixed for all edges at the square lattice isotropic critical value 
\be
\sinh{2K_c}=1. \label{crit}
\ee

Table \ref{table3} shows typical examples of the  free energies obtained, in  units of $K_B T$. The dual of the cube lattice was considered as a test to the stability of the numerical algorithms  for large lattice sizes. The partition function on a given planar lattice $Z$ and on the corresponding dual lattice $Z^*$ are related, at the critical point (\ref{crit}), by
\be
    \log{Z(K_c)}= \log{2}(N_V-\frac{N_E}{2}-1)+ \log{Z^*(K_c)}  \label{const}
\ee
with $(N_V-\frac{N_E}{2}-1)=1$ for the cube lattice. This condition is a non-trivial check on the precision of the numerical evaluation of the Pfaffians in Eq. (\ref{zz}). From the values given in table \ref{table3} we see that the condition (\ref{const}) is verified with a precision better than $1$ in $10^9$ and we can trust that the evaluation of the determinants of the adjacency matrices is not impaired by the large lattice sizes.

\begin{figure}[t]
\includegraphics[width=\figwidth]{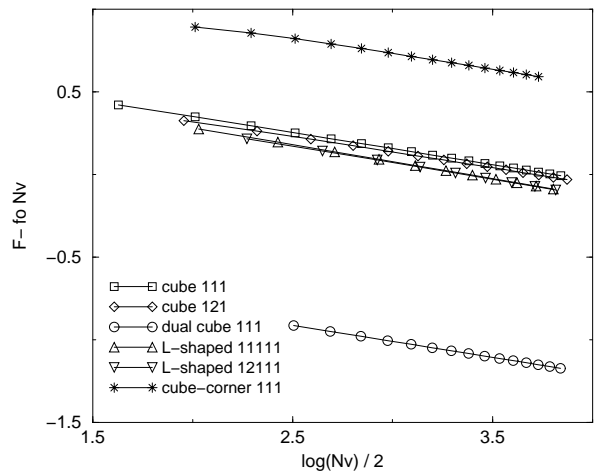}
\caption{The logarithmic correction for lattices with the topology of the sphere: the value of the residual free energy is given in function of the logarithm of the number of lattice vertices $N_V$. The uncertainties on the values are smaller than the symbols size.}
\label{fig4}
\end{figure}

\begin{table}[b] \centering
\begin{tabular}{|c|cc|c|cc|}\hline
lattice:& \multicolumn{2}{c|}{cube}& cube-corner& \multicolumn{2}{c|}{L-shaped} \\ \hline
shape:& (111)  & (121)&  (111)&  (11111) & (12111)\\ \hline
fits&$-$0.19468& $-$0.19043& $-$0.18675& $-$0.20531& $-$0.19827 \\
&$-$0.19465& $-$0.19100& $-$0.18938& $-$0.20566& $-$0.20150 \\
&$-$0.19463& $-$0.19142& $-$0.19093& $-$0.20573& $-$0.20275 \\
&$-$0.19461& $-$0.19175& $-$0.19190& $-$0.20572& $-$0.20338 \\
&$-$0.19459& $-$0.19201& $-$0.19254& $-$0.20571& $-$0.20376 \\
&$-$0.19458& $-$0.19222& $-$0.19298& $-$0.20569& $-$0.20402 \\
&$-$0.19456& $-$0.19240& $-$0.19329& $-$0.20567&    -    \\\hline 
expected: & \multicolumn{3}{c|}{$-$0.19444}& \multicolumn{2}{c|}{$-$0.20556}\\\hline
\end{tabular}
\caption{Successive fits to the logarithm correction $C$  for  lattices with the topology of the sphere and various shapes $(M_1,\ldots,M_N)$. Each value corresponds to the slope obtained by a linear regression using four consecutive free energy values. There is a smooth convergence to the predicted correction with increasing lattice size.}
\label{table1}
\end{table}

The logarithmic correction $C$ can be found  by subtracting the leading volume term $f_0$ to the free energy, that for a square lattice is \cite{ferdfish}
\be
    f_0= 2G/\pi + \frac{1}{2} \ln{2}
\ee  
where $G$ is the Catalan constant. 

In Fig. \ref{fig4}  the residual free energy $F - f_0 N_V$ is plotted as function of the logarithm of the lattice size, $\log{N_V}/2$, for various lattice shapes. We find a simple linear behavior and the coefficient $C$ is the slope of the curve in the limit of large lattice size. The factor of $1/2$ follows from (\ref{fexp}), we want the the logarithmic correction on the length scale $\sqrt{N_V}$ and not on the lattice area $N_V$.

We evaluate a series of values converging to $C$  by doing linear regressions on sets of four consecutive lattices with increasing sizes. 
The values  obtained for the three lattices, with various shapes, are shown in table \ref{table1}. We find a clear convergence to a value of $C$, that is the sum over  the lattice conical singularities of the $C_\theta$ correction in Eq. (\ref{c2}).  The rate of convergence depends on the shape or aspect-ratios of the lattice, being fastest for the more symmetric lattices.

The conical singularities are of two kinds: vertex singularities for the cube and L-shaped lattices and face singularities for the dual cube lattice. This distinction is artificial since the two types are related by a duality transformation. We can therefore consider only vertex singularities, with three and five  square faces meeting at a single vertex,  corresponding  respectively to $\theta$ angles of $3\pi/2$ and $ 5\pi/2$. The corresponding correction $C_\theta$ are listed on Table \ref{table0}. The expected logarithmic correction $C$ for a given lattice, see table  \ref{table1}, is obtained by summing  the $C_\theta$ for all singularities occurring in that lattice.

\begin{figure}[t]
\includegraphics[width=0.5\linewidth]{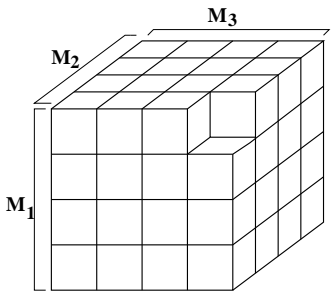}
\caption{The cube lattice with a corner removed. The new corner has six vertices where three or five squared faces meet but the conical singularity remains the same. Regardless of the detailed  corner structure, both lattices at large scale fold into a $\theta=3\pi/2$ conical shape around each corner.}
\label{fig44}
\end{figure}

The nature of the conical singularities demands further clarification, since it is not immediately clear how to related the combinatorics of the lattice structure with continuum concepts as  conical singularities in the metric of a surface. For the purposes  considered in this paper, namely finite size corrections to the free energy, the important aspect to obtain a conical singularity behavior seems to be the large scale structure of the lattice and not the detailed small scale structure.

\begin{table}[b] \centering
\begin{tabular}{|c|ccc|}\hline
\# squares  &   $\epsilon$    &   $\theta$    &  $C_\theta$\\\hline
   3  &     $\pi/2$  &  $3 \pi/2$   & $-$0.024306\\
   5  &    $-\pi/2$&    $5 \pi/2$  &   0.01875\\
   8  &    $-2\pi$ &    $4 \pi $   &  0.0625\\\hline 
\end{tabular}
\caption{The deficit angle $\epsilon$, the spanning angle $\theta$ and the corresponding $C_\theta$ correction for conical singularities with three, five and eight squared faces meeting at a vertex or face.}
\label{table0}
\end{table}

Consider for instance the cube lattice with a corner removed, as shown in Fig. \ref{fig44}. In the cube lattice  three squared faces meet at each corner. In the lattice of  Fig. \ref{fig44} one of the corners has three vertices where three squared faces meet and three vertices where five squared faces meet. Regardless of these local differences, at large scale the corner for both lattices can be seen as a sheet of flat square lattice folded into a $\theta=3\pi/2$ conical shape. It is this fact that determines the conical singularity behavior and not the detailed lattice combinatorics.

That this is the case can be seen from Fig. \ref{fig4} and table \ref{table1} where it is shown that the cube lattice with the corner removed has the same logarithmic correction $C$ as the simple cube lattice. The different rate of convergence can be seen as due to the  lattice defects introduced by the changes in the corner.

\section{Higher genus lattices: the torus and genus two} \label{section3}

We now  consider the genus two lattice shown in  Fig. \ref{fig2}, the toroidal lattice, whose  logarithmic correction is zero, will also be studied as a non-trivial check of the methods used.

For a  genus $g$ lattice the partition function is expressed in terms of the Pfaffians of $4^g$ adjacency matrices, four for the torus and sixteen for the genus two lattice. Some  of these Pfaffian vanish at criticality in the thermodynamic limit, while the remaining ones will give the same contribution to the logarithmic correction. This follows from the fact \cite{ferdfish,me1} that the various Pfaffians converge rapidly to a common bulk term times a topology and shape determined factor,
\be
       \Pf A_i (N_V)= \Theta_i \ \tilde{Z}(N_V)
\ee
where the $ \Theta_i$ are constants for large enough lattice size $N_V$. The free energy $F$ follows from (\ref{zz}) to be 
\be
   - F=  N_v \ln{2} +  N_e \ln{ \cosh{K}} + \ln{\left(\frac{1}{2^g} \sum_{i=1}^{4^g} \alpha_i \Theta_i \right)} + \ln{\tilde{Z}}.
\ee
The logarithmic correction is entirely due to the $\tilde{Z}$ term and can be equivalently evaluated from any non-vanishing Pfaffian. For this purpose it is convenient to introduce the auxiliary quantities
\be
   F_i(N_V)= F(N_V) - \ln{\Theta_i} +  \ln{\left(\frac{1}{2^g} \sum_{k=1}^{4^g} \alpha_k \Theta_k \right)} 
\ee
where $i$ is a label of the $4^g$ un-equivalent Kasteleyn orientations. In the thermodynamic limit  the $F_i$ are distinct from the free energy $F$ by a constant only.

For both lattices we evaluate the determinants of the adjacency matrices by keeping the lattice aspect ratios $m_i$ fixed and increasing the lattice size according to a scale $L$, with the lattice dimensions being $M_i=m_i L$. Examples of the evaluated $F_i$, for different Kasteleyn orientations, are shown in table \ref{table3} for various lattice sizes.

\begin{figure}[t]
\includegraphics[width=\figwidth]{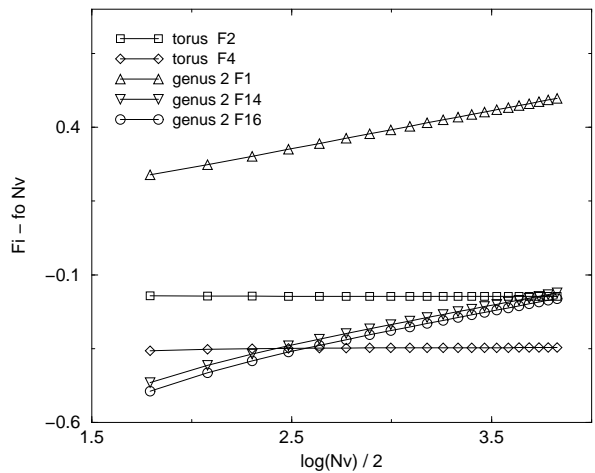}
\caption{The logarithmic correction for the torus and the genus two lattice: the value of the residual free energy is plotted in function of the logarithm of the number of lattice vertices $N_V$. For the torus and the  genus two lattice, respectively two and three  un-equivalent adjacency matrices are considered.}
\label{fig5}
\end{figure}

\begin{table} \centering
\begin{tabular}{|c|cc|cc|}\hline
lattice:& \multicolumn{2}{c|}{torus (11)}&  \multicolumn{2}{c|}{genus 2 (11111)} \\ \hline
Pfaffian:&  (+ $-$)  & (+ +)   &  ($-$ $-$ + $-$)   & ($-$ $-$ $-$ $-$)   \\ \hline
fits&$-$0.00020&  0.00080 &  0.12858   &   0.12942  \\
&$-$0.00017&  0.00070 &  0.12819  &   0.12895  \\
&$-$0.00016&  0.00062 &  0.12786  &   0.12855  \\
&$-$0.00014&  0.00056 &  0.12759   &   0.12822  \\
&$-$0.00012&  0.00050 &  0.12735  &   0.12793  \\
&$-$0.00011&  0.00045 &  0.12715  &   0.12768  \\
&$-$0.00010&  0.00041 &  0.12698   &   0.12747  \\\hline
expected: & \multicolumn{2}{c|}{0}& \multicolumn{2}{c|}{ 0.125 }\\\hline
\end{tabular}
\caption{Successive fits to the logarithm correction $C$ for the toroidal and genus two lattices with shapes $(M_1,\ldots,M_N)$, using different Kasteleyn orientations. The Kasteleyn orientations are denoted by $\pm$ signs labeling the parity along the lattice cycles with non-trivial topology.}
\label{table2}
\end{table}

Subtracting the leading volume contribution to the free energy, the residual free energy $F-f_0 N_V$ is found to have a linear behavior on the logarithm of the lattice size $\log{N_V}/2$, see Fig. \ref{fig5}. The coefficient $C$ is the slope of these curves in the thermodynamic limit. As in the previous section we generate a sequence of values converging to $C$ by doing linear regressions on sets of four lattices with consecutive lattice sizes. 
The results obtained are shown in table \ref{table2}. 

As expected the torus behaves as a flat lattice, while the genus two lattice reproduces the correction (\ref{c2}) expected for two conical singularities with deficit angle $\epsilon=-2\pi$. These singularities correspond to the two octagonal lattice faces, shown in dashed line in Fig. \ref{fig2}, that in a dual description are vertices where eight squares meet.

The independence of the results on the choice of the Kasteleyn orientation is observed in  Fig. \ref{fig5}, curves  for different $F_i$ show equal slopes. The vertical displacement is due to different $\log \Theta_i$ among the Pfaffians of the different adjacency matrices.

For the toroidal lattice the coefficient $C$ vanishes at a rate consistent with the rates of convergence observed for $C$ in the cube and L-shaped lattices. The convergence is somewhat slower for the genus two lattice, for which the rate is similar to the ones observed on the less symmetric genus zero lattices. For all cases the validity of Eq. (\ref{c2})  is established without doubt.

\section{Conclusions} \label{section4}

In this paper we have provided an answer to the question, posed by Cardy and Peschel in \cite{cardy2}, to which of the two forms (\ref{c1}) or (\ref{c2}) for the logarithmic correction to the free energy  is observed in specific lattice models. 
We found that for regular lattices with conical singularities the logarithmic correction $C$ is given  by Eq. (\ref{c2}). This was observed  for a number of lattice boundary conditions, with various lattice topologies and shapes, provided that the lattice is locally a flat regular lattice with the exception of a few  vertices where  at large scale it can be seen as folded into a conical shape.

An interesting question that remains to be answered is if there are lattice realizations of Eq. (\ref{c1}), the smooth metric contribution. 

\bigskip
\centerline{{\bf Acknowledgments}}
\bigskip

 This work was partially supported  by the EU grant HPRN-CT-1999-00161 and by the National Science Foundation (USA) under Grant No. DMR-0073058. The author profited from many useful discussions with prof. Barry McCoy.

\begin{table} \centering
\begin{tabular}{|c|cc|c|cc|}\hline
& \multicolumn{2}{c|}{$- F$ for cube lattice } & &\multicolumn{2}{c|}{$-F_i$ for genus 2 lattice} \\ \hline
L & (1,1,1) &  dual (1,1,1)     & L &   ($-$ $-$ + $-$)   & ($-$ $-$ $-$ $-$)  \\ \hline
 11&    559.558730&  558.865583 &  14&  729.106047 & 729.125841 \\
 12&    676.718852&  676.025704 &  15&  836.941813 & 836.961544 \\
 13&    805.033712&  804.340565 &  16&  952.215747 & 952.235424  \\
 14&    944.503569&  943.810422 &  17&  1074.92777 & 1074.94740  \\
 15&    1095.12862&  1094.43547 &  18&  1205.07782 & 1205.09741 \\
 16&    1256.90902&  1256.21588 &  19&  1342.66584 & 1342.68540 \\
 17&    1429.84491&  1429.15176 &  20&  1487.69179 & 1487.71132 \\
 18&    1613.93638&  1613.24323 &  21&  1640.15563 & 1640.17514 \\
 19&    1809.18352&  1808.49038 &  22&  1800.05734 & 1800.07682 \\
 20&    2015.58641&  2014.89326 &  23&  1967.39687 & 1967.41633  \\\hline
 \end{tabular}
\caption{Free energy $F$ for the cube lattice and its dual with aspect ratios $(M_1,M_2,M_3)=(1,1,1)$ and the $F_i$ free energies for the genus two lattice with aspect ratio$ (M_k:k=1,\ldots,5)=(1,1,1,1,1)$, for two distinct Kasteleyn orientations.}
\label{table3}
\end{table}

\end{document}